\documentclass[aps,prl,amsmath,amssymb,amsfonts,twocolumn]{revtex4}
\usepackage{graphicx}
\usepackage{dcolumn}
\usepackage{bm}
\usepackage{amsmath}
\usepackage{amssymb}
\usepackage{mathcomp}
\usepackage{color}

\newcommand{\beq}{\begin{equation}}
\newcommand{\eeq}{\end{equation}}
\newcommand{\bea}{\begin{eqnarray}}
\newcommand{\eea}{\end{eqnarray}}

\newcommand{\R}{\mathbb{R}}

\begin{document}

\title{Flow equation holography}

\author{S. Kehrein}
\affiliation{Department of Physics, Georg-August-Universit{\"a}t G{\"o}ttingen,
Friedrich-Hund-Platz~1, 37077~G{\"o}ttingen, Germany}

\begin{abstract}
The Ryu-Takayanagi conjecture establishes a remarkable connection between quantum systems
and geometry. Specifically, it relates the entanglement entropy to minimal surfaces within the
setting of AdS/CFT correspondence. This Letter shows how this idea can be generalised to
generic quantum many-body systems within a perturbative expansion where the region whose 
entanglement properties one is interested in is weakly coupled to the rest of the system. A simple
expression is derived that relates a unitary disentangling flow in an emergent RG-like direction 
to the min-entropy of the region under consideration. Explicit calculations for critical free fermions
in one and two dimensions illustrate this relation. 
\end{abstract}

\maketitle

AdS/CFT correspondence \cite{Maldacena98,Gubser98} as an explicit realization of the holographic principle \cite{tHooft93,Susskind95}
has become a powerful tool for understanding properties of strongly interacting
quantum field theories. It not only establishes a connection between certain classes of gauge theories in
the \textquoteright t~Hooft limit and classical theories of gravity with an additional emergent RG-like dimension, 
but also led to a new way of thinking about quantum field theories. For example, the Ryu-Takayanagi conjecture 
\cite{RyuTakayanagi06a,RyuTakayanagi06b,Casini11}
demonstrates a remarkable relation between the entanglement entropy of a region~$A\subset\R^d$ in a $d+1$~dimensional 
conformal field theory and the static minimal surface in the dual $d+2$~dimensional AdS-space with a boundary given
by~$\partial A$. This is important both conceptually and also from a practical point of view since the calculation
of entanglement entropies is challenging even for ground states of simple systems, with few analytical or numerical 
tools available, especially for~$d>1$. It has become increasingly clear in the past decade that the
entanglement entropy plays a fundamental role in fields as diverse as quantum information theory \cite{NielsenChuang}, 
the efficient simulation of quantum systems \cite{RieraLatorre06,Eisert10} or topological phases of matter \cite{KitaevPreskill06}.

Therefore the insight described by the Ryu-Takayanagi conjecture spurred a lot of activity to find similar relations between 
entanglement properties in $d+1$~dimensions and geometry in $d+2$~dimensions beyond the limitation to AdS/CFT-theories.
Especially the geometric picture underlying the multi-scale entanglement renormalization ansatz (MERA) \cite{Vidal07} and 
its continuous version cMERA \cite{Haegeman13} bears strong resemblance to AdS/CFT-correspondence and 
the Ryu-Takayanagi conjecture \cite{Swingle12}, while in principle being applicable for a wide class of quantum many-body systems.
Still it remains to be explored to what extent these tools can also be used for a quantitative calculation of entanglement properties. 

\begin{figure}[t] 
\includegraphics[width=0.6\linewidth]{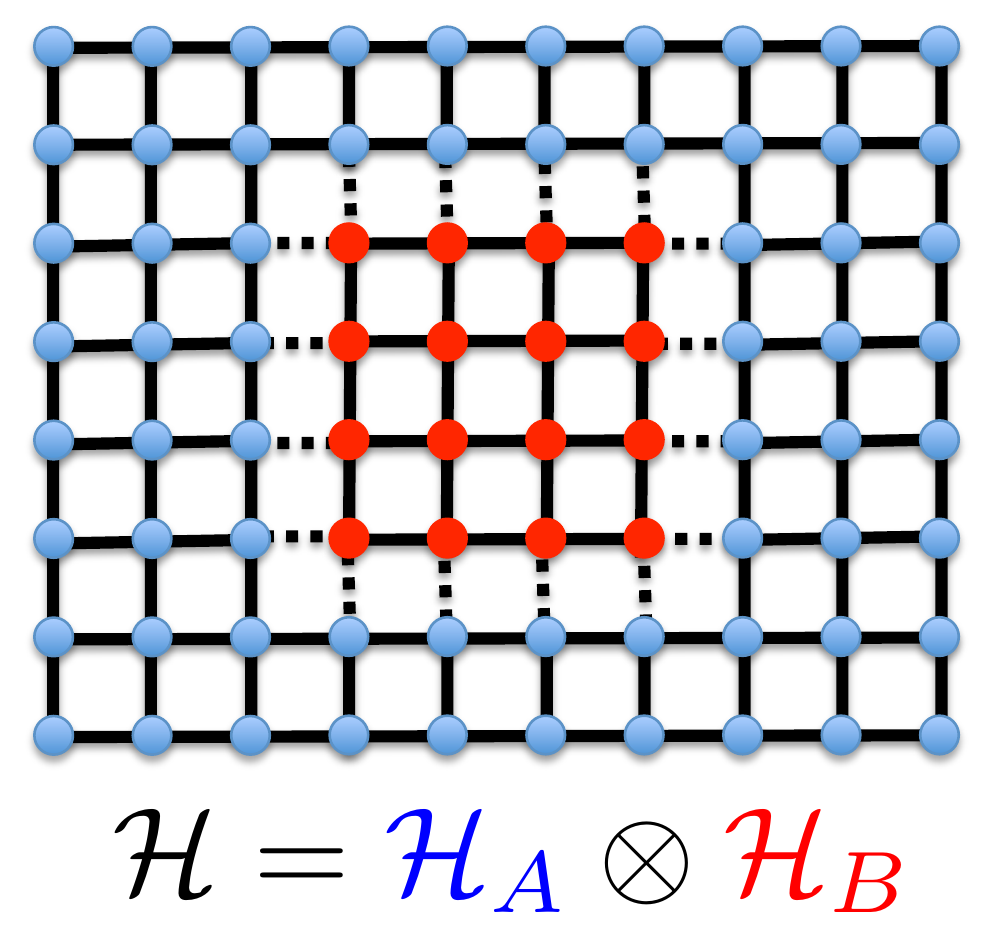}
\caption{\label{fig_weaklink}
The weak-link limit for a lattice model: We are interested in the min-entropy of subregion~$B$ (red circles)
which is weakly coupled (depicted by broken lines) to the rest of the system (blue circles). The relative
strength of couplings between regions $A$ and $B$ vs.\ the couplings within $A$ or $B$ determines the expansion parameter~$g$.
}
\end{figure}

Motivated by these ideas this Letter develops a systematic procedure which connects the entanglement properties
of eigenstates of a generic quantum many-body Hamiltonian to a disentangling flow in an emergent RG-like dimension. 
This is done by working in a controlled limit where the subregion~$B$ whose entanglement we are interested in is
weakly coupled to the rest of the system, see Fig.~\ref{fig_weaklink}. 
I denote the relative strength of this coupling by~$g$ with $|g|\ll 1$ being the
weak-link limit. $g=1$ will correspond to translation-invariant systems in the models discussed explicitly later on.

For a bipartite Hilbert space ${\cal H}={\cal H}_A\otimes {\cal H}_B$ the entanglement of region~$B$ 
for a pure state $|\psi\rangle\in {\cal H}$ is described by its R\'enyi entropies
\beq
S^{(q)} = \frac{1}{1-q} \ln {\rm tr}_{{\cal H}_B}\,\rho_B^q \ ,
\eeq
where $q$ is a positive real number and $\rho_B={\rm tr}_{{\cal H}_A}\,|\psi\rangle\langle\psi|$ is the reduced
density matrix of subsystem~$B$. The limit~$q\rightarrow 1$ yields the von~Neumann entanglement entropy.
The challenge for the analytic calculation of entanglement entropies of gapless models (and the reason why relatively little is known 
analytically even in one dimension apart from free theories or conformal field theories) is that the eigenvalues of
the reduced density matrix behave non-perturbatively with the size of region~$B$. This is even true in the weak-link limit
and one also finds non-perturbative behavior in~$g$ \cite{Eisler10}.

As worked out below, this problem can be resolved for the R\'enyi entropy in the limit $q\rightarrow\infty$, which is also
called the min-entropy or single-copy entropy \cite{Eisert10}. The min-entropy is only determined by the
largest eigenvalue~$p_{\rm max}$ of the reduced density matrix~$\rho_B$,
\beq
S^{\rm (min)}=\lim_{q\rightarrow\infty} S^{(q)}=-\ln p_{\rm max} \ .
\eeq
We will see that $p_{\rm max}$ has a specific product structure emerging from the disentangling flow 
that leads to a non-perturbative result for $p_{\rm max}$ even if the individual factors are only known perturbatively in~$g$.
The goal of this Letter is to give an explicit expression for $S^{\rm (min)}$ which is correct in order~$g^2$ based on
a disentangling flow without making strong AdS/CFT-like restrictions on the class of permissible systems. 
The focus on $S^{\rm (min)}$ as opposed to the von~Neumann entropy is also one main difference to prior
work on disentangling flows \cite{Nozaki12}, which allows me to give an analytic justification 
in the weak-link limit. It should also be mentioned that the min-entropy has an important quantum information
theory interpretation, namely as the amount of distillable entanglement from a single copy of the quantum state~$|\psi\rangle$,
as opposed to the von~Neumann entropy which requires many identically prepared quantum states \cite{Eisert10}.

The key step in my approach is the construction of an RG-like disentangling flow described by unitary transformations $U(u)$.
In this Letter we are interested in the entanglement properties of eigenstates of Hamiltonians, $H|\psi\rangle=E|\psi\rangle$.
In order to find a disentangling flow for $|\psi\rangle$ we unitarily transform the Hamiltonian 
$H(u)\stackrel{\rm def}{=} U(u)\,H\,U^\dagger(u)$ in such a way that $H(u=\infty)$ only contains terms which act locally
in ${\cal H}_A$ or ${\cal H}_B$, and no longer contains terms which generate entanglement across the links 
between regions $A$ and~$B$. Since the unitarily transformed state $|\psi(u)\rangle=U(u)|\psi\rangle$ is always
an eigenstate of~$H(u)$, this implies that $|\psi(u=\infty)\rangle$ must be a disentangled state since it is an
eigenstate of~$H(u=\infty)$. 

A unitary flow can be obtained following the flow equation method \cite{Wegner94,Kehrein06} 
which generates $U(u)$ from a sequence of infinitesimal unitary transformations with an antihermitean generator~$\eta(u)$
\beq
\frac{dH(u)}{du}=[\eta(u),H(u)] 
\label{flow_H}
\eeq
and initial condition $H(u=0)=H$. The unitary transformation is obtained as an $u$-ordered exponential
\beq
U(u) = T_u \exp\left(\int_0^u du'\,\eta(u')\right) 
\eeq
where the larger values of~$u'$ are ordered to the left. A flow with the desired properties is obtained by splitting up
the Hamiltonian as
\beq
H(u)=H_A(u)+H_B(u)+H_{\rm ent}(u)
\label{Hent}
\eeq
where only $H_{\rm ent}(u)$ contains terms that create entanglement by linking regions $A$ and~$B$. The generator is defined as
\beq
\eta(u)\stackrel{\rm def}{=} 2\Lambda^{-2}e^{2u}\,[H(u),H_{\rm ent}(u)]
\label{def_eta}
\eeq
where $\Lambda$ is the ultraviolet cutoff. One can show that under very general 
conditions this choice of the generator leads to a contracting flow, $\frac{d}{du} {\rm tr} (H^2_{\rm ent}(u))\leq 0$
and $\eta(u=\infty)=0$ \cite{Wegner94,Kehrein06}. While this does not guarantee that $H_{\rm ent}(u=\infty)$ 
vanishes, it commutes with $H_A$ and $H_B$ and does not contribute to entanglement between 
regions $A$ and~$B$ anymore (in the sense of not affecting the leading behavior of the entanglement 
entropy with the size of~$B$).

Previous applications of the flow equation method were such
that an interaction part in the Hamiltonian was eliminated by a sequence of infinitesimal unitary transformations
(see e.~g. Refs.~\cite{Wegner94,Kehrein06,Kehrein97,Kehrein99} and further references therein). The main
difference in this Letter is that I use flow equations to disentangle a Hamiltonian, that is to eliminate couplings 
between two regions for flow parameter $u\rightarrow\infty$. 
In practise Eqs. (\ref{flow_H}) and (\ref{def_eta}) lead to coupled systems of ordinary differential equations,
which for interacting many-body systems generically need to be truncated in order to render the hierarchy finite. 
This can be done easily in the weak-link limit since by definition $H_{\rm ent}$ is proportional to~$g$,
therefore the flow equations (\ref{flow_H}) and (\ref{def_eta}) can be organised in powers of~$g$. 
The resulting systems of differential equations can be solved either numerically or (usually approximately) analytically
(more below).

Notice that $u$ can be thought of as a logarithmic 
energy scale, $u=\ln\left(\Lambda/E\right)$, and that the generator $\eta(u)$ is dimensionless \cite{footnote_uB}.
Matrix elements in $H_{\rm ent}$ that couple states with an energy difference~$E$ are
unitarily transformed away at flow scale~$u$. So large energy differences are eliminated first (for small $u$) 
and smaller energy differences later along the flow, which makes the flow equation method RG-like. 
Energy differences correspond to length scales and for translation-invariant systems (meaning: translation-invariant
apart from the weak links) the unitary transformation acts on sites that are further and further apart from the boundary
between regions $A$ and~$B$ as $u$ increases. This means that the Hilbert spaces and the unitary transformation can be factorized
\beq
{\cal H}_{A,B}=\bigotimes_i {\cal H}^{(i)}_{A,B} \ , \quad U(u)=\prod_i U^{(i)}(u)
\eeq
such that $U^{(i)}$ only acts nontrivially on ${\cal H}^{(i)}_A \otimes {\cal H}^{(i)}_B$. The index~$i$ does not need to refer
to single sites, ${\cal H}^{(i)}_{A,B}$ can contain many sites as long as a perturbative argument for $U^{(i)}$ remains valid (see below).

\begin{figure}[tb] 
\includegraphics[width=1.0\linewidth]{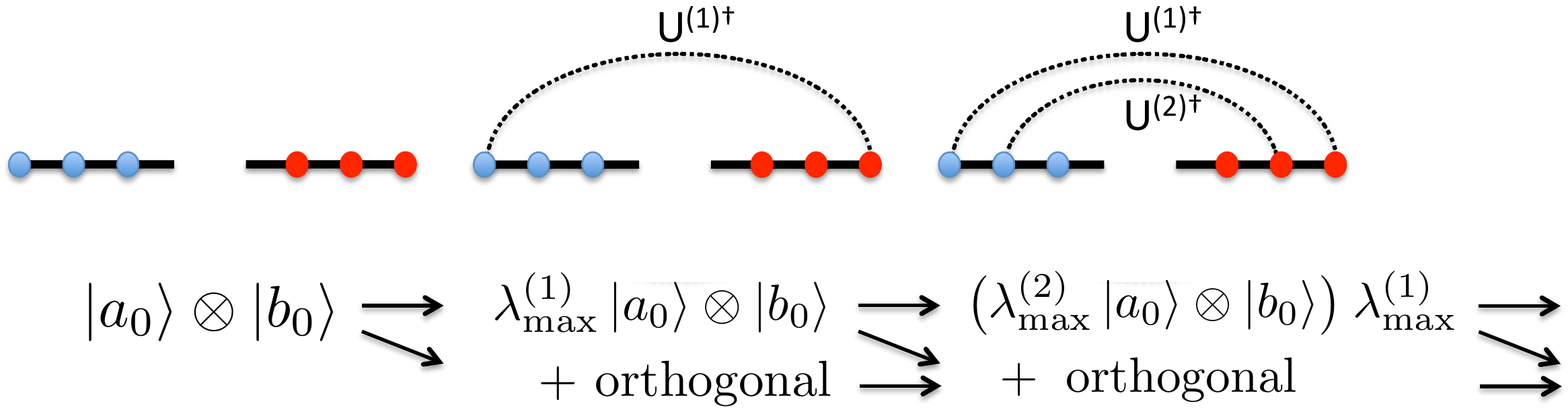}
\caption{\label{fig_schmitt}
Product structure for the largest Schmitt coefficient of the state $|\psi\rangle\in {\cal H}_A\otimes {\cal H}_B$, see text.
}
\end{figure}

If one reconstructs $|\psi\rangle$ from the disentangled state $|a_0\rangle\otimes |b_0\rangle=U(\infty)|\psi\rangle$, 
one therefore has a 
structure as depicted in Fig.~\ref{fig_schmitt}. Each individual $U^{(i)}(\infty)$ is close to the identity in the weak-link limit,
hence the component $|a_0\rangle\otimes |b_0\rangle$ of the total wavefunction gets multiplied by a factor 
$\lambda^{(i)}_{\rm max}=1-O(g^2)$ under the action of $U^{(i)\dagger}(\infty)$. In addition, $U^{(i)\dagger}(\infty)$ generates a component
orthogonal to $|a_0\rangle\otimes |b_0\rangle$, which however cannot contribute to the coefficient in front of 
$|a_0\rangle\otimes |b_0\rangle$ anymore in later steps. This implies that the Schmitt decomposition of~$|\psi\rangle$
takes the form 
$
|\psi\rangle=\sum_{j} \lambda_j\, |a_j\rangle\otimes |b_j\rangle
$
with the largest Schmitt coefficient $\lambda_0=\prod_i \lambda^{(i)}_{\rm max}$. We will show
$\lambda^{(i)}_{\rm max}=1-\alpha^{(i)}g^2$ with coefficients $\alpha^{(i)}$ that can be calculated perturbatively,
therefore 
\bea
S^{\rm (min)} &=&-\ln p_{\rm max}= -\ln \lambda_0^2 = -2 \ln \prod_i \lambda^{(i)}_{\rm max} \nonumber \\
&=& -2 \sum_i \ln \lambda^{(i)}_{\rm max} 
= 2g^2 \sum_i \alpha^{(i)} +O(g^4) \ .
\label{Smin_1}
\eea
In this manner one avoids calculating $p_{\rm max}$ with its non-perturbative behavior by making use of the product
structure for the largest Schmitt coefficient depicted in Fig.~\ref{fig_schmitt}. It is sufficient to find the individual
$\lambda^{(i)}_{\rm max}$ perturbatively in the weak-link limit. This is a straightforward exercise in second order
perturbation theory: If one weakly perturbs a Hamiltonian $H_0$, $H_g=H_0+g\,H_{\rm int}$, then the overlap
of the perturbed eigenstate with the unperturbed eigenstate can be written as
\beq
|\langle\psi_g|\psi_0\rangle|=1-\int_0^\infty du\left(-\langle\psi_0|\eta^2(u)|\psi_0\rangle\right) +O(g^3) 
\eeq
with the flow generated from (\ref{flow_H})
and $\eta(u)\stackrel{\rm def}{=} 2\Lambda^{-2}e^{2u}\,[H_g(u),g\,H_{\rm int}(u)]$ like in (\ref{def_eta}). This
implies for~(\ref{Smin_1})
\beq
S^{\rm (min)} = -2 \sum_i \int_0^\infty du\, \langle\psi|(\eta^{(i)}(u))^2|\psi\rangle +O(g^3) 
\eeq
where $\eta^{(i)}$ is the infinitesimal generator corresponding to~$U^{(i)}$. Since by construction
$\langle\psi| \eta^{(i)}(u)\,\eta^{(j)}(u)|\psi\rangle\propto \delta_{ij}$ we can define
\beq
S^{\rm (feq)} \stackrel{\rm def}{=} -2 \int_0^\infty du\,\langle\psi|\eta^2(u)|\psi\rangle
\label{Sfeq}
\eeq
and have shown
\beq
S^{\rm (min)}=S^{\rm (feq)} +O(g^3) \ .
\label{SminSfeq}
\eeq
These two formulas constitute the main result of this Letter. They connect the min-entropy to the integral over 
the flow equation disentangling density 
measured by $D(u)\stackrel{\rm def}{=}-\langle\psi|\eta^2(u)|\psi\rangle$ on a logarithmic energy scale ($D(u)\geq 0$ since 
$\eta(u)$ is antihermitean).

\begin{figure}[tb] 
\includegraphics[width=1.0\linewidth]{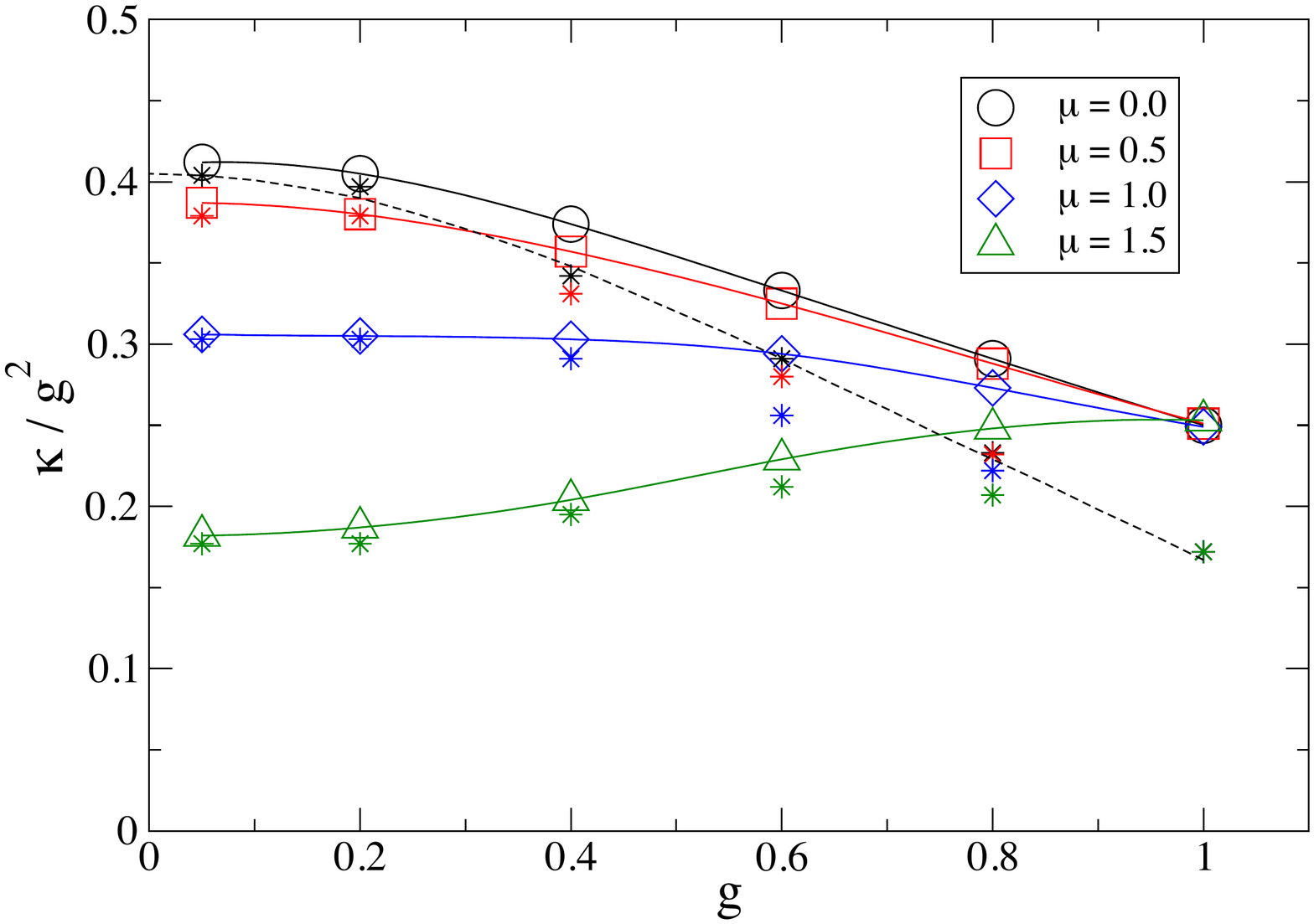}
\caption{\label{fig1d}
Behavior of the min-entropy (\ref{kappa_g}) for the ground state of $1d$~free fermions (\ref{def_Hd1}) as a function of the
link strength $g$ for different chemical potentials~$\mu$. $\mu=0$ corresponds to half filling, all results are
symmetric around half filling (only positive values of $\mu$ are shown) and the bandwidth is $[-2,2]$. 
The big symbols (circles, squares, diamonds and triangles) 
depict flow equation holography results from (\ref{Sfeq}) with the full lines being guides to the eye. 
Stars show the corresponding results from exact diagonalization. All data points have a relative error of about~$3\%$.
The dashed line is the exact analytical result for the min-entropy at half filling from Ref.~\cite{Eisler10}.
}
\end{figure}

In order to demonstrate this approach we study critical free fermions in $d=1$ and $d=2$ explicitly. 
This will permit us to compare with various exact analytical or well-controlled numerical results. In $d=1$ we consider
the Hamiltonian
\beq
H=-\sum_{i=-\infty}^\infty t_i \, \left(c^\dagger_i c_{i+1} +{\rm h.c.}\right)
\label{def_Hd1}
\eeq
with $t_0=t_\ell=g$ and $\forall i\neq 0,\ell \;\; t_i=1$. We are interested in the ground state entanglement entropy of 
the lattice sites $\{1,\ldots ,\ell\}$ with the rest of the chain. For the translation-invariant case $g=1$ the behavior
is logarithmic in~$\ell$ with a universal prefactor \cite{Holzhey94,Calabrese04,Calabrese09} independent
of the filling,
$
S^{(q)}=\frac{1+q^{-1}}{6} \ln\ell +O(\ell^0)
$.
Our goal is now to extract this prefactor in the min-entropy as a function of the link strength~$g$ 
\beq
S^{\rm (min)}=\kappa(g)\,\ln\ell+O(\ell^0) 
\label{kappa_g}
\eeq
using the disentangling flow.  We do this by considering (\ref{def_Hd1}) with one weak link 
($t_0=g$ and $\forall i\neq 0\;\; t_i=1$) and following the disentangling flow separating the 
left half line from the right half line. Initially $H_{\rm ent}(u=0)=-g(c^\dagger_0 c_1+{\rm h.c.})$ and
during the flow $H_{\rm ent}(u)$ in (\ref{Hent}) is defined to consist of the hopping terms in $H(u)$ connecting the
sites $\{-\infty,\ldots ,0\}$ to the sites $\{1,\ldots ,\infty\}$. The resulting system of differential equations (\ref{flow_H})
is solved numerically (without approximations or truncations). After an initial transient one finds 
that the flow equation disentangling density $D(u)$ becomes constant. An upper cutoff
in (\ref{Sfeq}) is set by the observation that $H_{\rm ent}(u)$ becomes increasingly non-local during the flow: The initial
nearest neighbor hopping in $H_{\rm ent}(u=0)$ turns into hopping terms between
sites with separation $d$ (compare Fig.~\ref{fig_schmitt}) at the scale $u=\ln d$. Since we are interested in 
decoupling an interval of length~$\ell$ in (\ref{def_Hd1}) this implies we should only integrate up to values
$u^*=\ln\ell$ in (\ref{Sfeq}). Taking into account that the entanglement contributions from the two boundaries in (\ref{def_Hd1})
add up (for sufficiently large~$\ell$) this gives $\kappa(g)= 4D$. Notice that a proportionality factor~$p$ in
$u^*=\ln(p\,\ell)$ only contributes to the $O(l^0)$-term in (\ref{kappa_g}) and does not affect the logarithmic behavior.

The results are shown in Fig.~\ref{fig1d} as a function of the link strength~$g$ and for various fillings
($\kappa(g)/ g^2$ is depicted since $\kappa(g)\propto g^2$ for small~$g$). As expected 
from (\ref{SminSfeq}) the flow equation
results agree with exact diagonalization results for the min-entropy for small~$g$ \cite{footnote_check}. 
Not surprisingly there are differences between $S^{\rm (min)}$ and $S^{\rm (feq)}$ for larger values of~$g$.
For the translation-invariant case~$g=1$ the exact diagonalization results become independent from the filling
as known from universality with $\kappa=1/6$. Interestingly, within numerical accuracy the logarithmic prefactor in the 
flow equation expression $S^{\rm (feq)}$ also becomes
independent from filling at this point. 

\begin{figure}[tb] 
\includegraphics[width=1.0\linewidth]{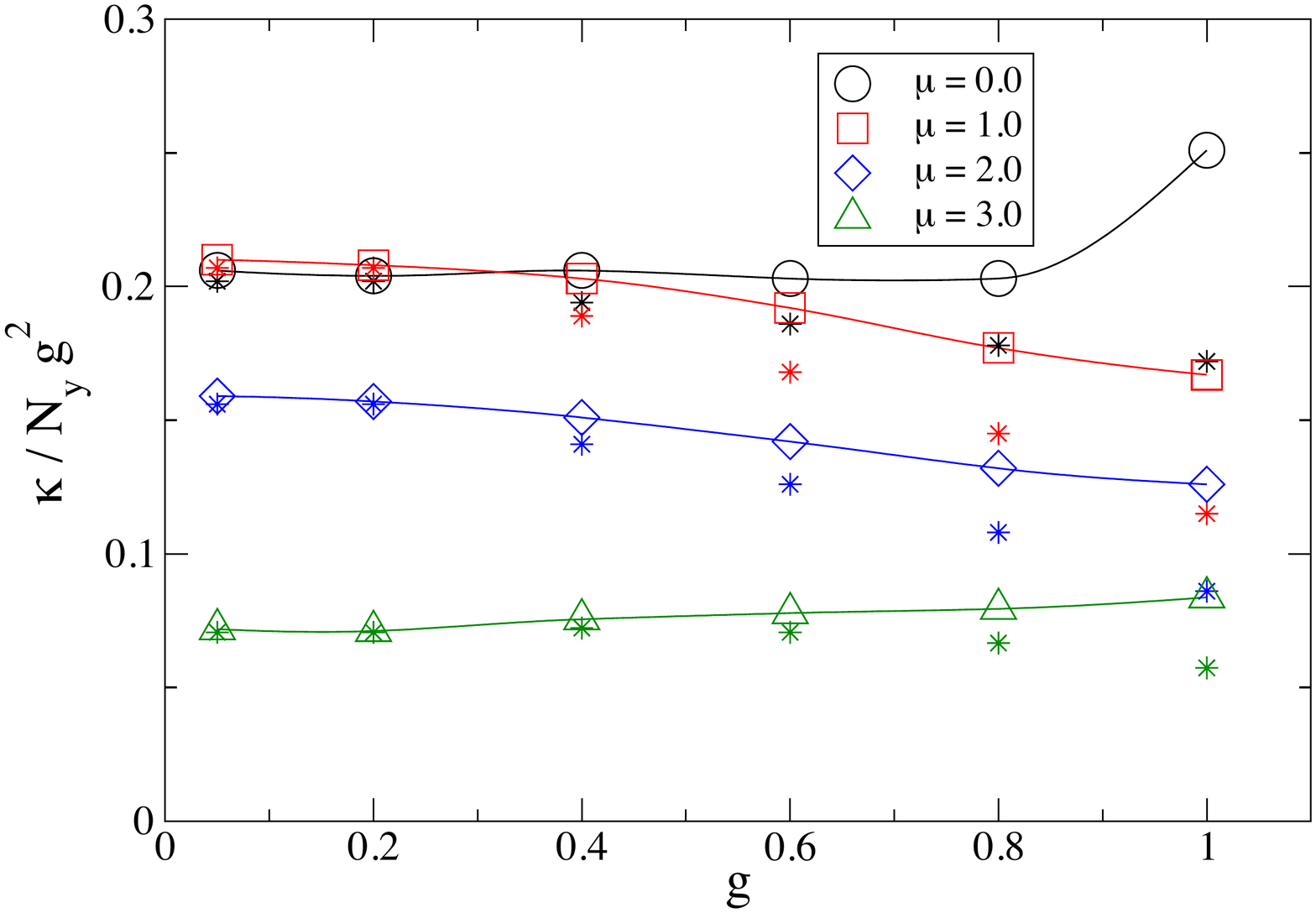}
\caption{\label{fig2d}
Behavior of the min-entropy (\ref{kappa_g}) for $2d$~free fermions on a square lattice as a function of the
link strength $g$ for different chemical potentials~$\mu$. $S^{\rm min}$ is calculated for stripes of dimension $\ell\times N_y$,
see text. $\mu=0$ corresponds to half filling, all results are
symmetric around half filling (only positive values of $\mu$ are shown) and the bandwidth is $[-4,4]$. 
The big symbols (circles, squares, diamonds and triangles) 
depict flow equation holography results from (\ref{Sfeq}) with the full lines being guides to the eye. 
Stars show the corresponding results from exact diagonalization. All data points have a relative error of about~$3\%$.
}
\end{figure}

Next we test the flow equation disentangling approach for a two-dimensional system, namely for the ground state of 
free fermions with nearest neighbor hopping ($t=1$) on a square lattice which is infinite in the $x$-direction and 
has periodic boundary conditions in the $y$-direction (circumference $N_y\gg 1$ sites). 
We are interested in the min-entropy of a stripe with dimensions $\ell\times N_y$, which is coupled
to the rest of the system with hopping matrix elements~$g$. The general behavior is an area law with logarithmic 
corrections \cite{Eisert10}, that is $\kappa(g)\propto N_y$ in (\ref{kappa_g}). This comes out naturally
from (\ref{Sfeq}) since $-\langle\psi| \eta^2(u) |\psi\rangle\propto N_y$ for large~$N_y$. Results from the numerical solution
of the flow equations are depicted in Fig.~\ref{fig2d} and compared with exact diagonalization \cite{footnote_ED2d}. For small~$g$ one
again finds perfect agreement as expected from (\ref{SminSfeq}). 

In the one-dimensional case one can also interpret (\ref{Sfeq}) from the point of view of minimal curves 
like in the Ryu-Takayanagi conjecture \cite{RyuTakayanagi06a,RyuTakayanagi06b}. With the emergent
dimension $z=\Lambda^{-1}\,e^u$ one can verify that the leading behavior of (\ref{Sfeq}) can also be written as
the length of the minimal curve, $S^{\rm (feq)}={\rm min}_{\cal C} L({\cal C})$, for given UV-boundary conditions
$\partial {\cal C}=\partial B$ with respect to the metric
\beq
ds^2=\frac{4D\left(\ln(z\Lambda)\right)^2\,dz^2+dx^2}{z^2} \ .
\label{metric}
\eeq
For the critical fermions discussed before this is just an AdS${}_2$-metric since the flow equation disentangling density
is constant on a logarithmic scale, $D(u)={\rm const.}$
The spatial part $z^{-2}\,dx^2$ in (\ref{metric}) effectively describes the increasing non-locality of $H_{\rm ent}(u)$ 
during the flow and thereby introduces the IR-cutoff~$u^*$ once the length scale generated by the disentangling
flow exceeds the size of the interval~$B$ (here assuming that the total system is infinite). 
The minimal curve condition in the Ryu-Takayanagi conjecture finds a natural
explanation in the flow equation framework since IR-cutoffs in the disentangling flow at a given boundary point
are naturally set by the next boundary point closest to it. Consider for example a tight-binding chain (\ref{def_Hd1}) consisting of
$N_x\gg 1$ lattice sites with periodic boundary conditions split up in two intervals with $\ell_A$ and $\ell_B$ sites,
$\ell_A+\ell_B=N_x$. Then (\ref{Sfeq}) gives 
\beq
S^{\rm (feq)}=\kappa(g)\ln {\rm min}(\ell_A,\ell_B)+O(\ell^0) \ ,
\eeq
which agrees with the minimal curve prescription using (\ref{metric}) and the known exact behavior \cite{footnote_disjoint}. 

In summary, I have introduced a method for calculating entanglement entropies for eigenstates of generic many-body 
Hamiltonians from a disentangling flow in an emergent RG-like direction (\ref{Sfeq}). In the weak-link limit we have seen 
(\ref{SminSfeq}) that this procedure gives the min-entropy. While the specific examples discussed in this Letter were 
ground states of free fermion systems, the method also works for excited states and interacting systems. Notice that results in the
weak-link limit can also be obtained analytically, although I have consistently used numerical methods for the
solution of (\ref{flow_H}) in order to show results for all link strengths. 

I acknowledge valuable discussions with P.~Calabrese, B.~Dagallier and H.~Eisenlohr.


\begin{thebibliography}{99}

\bibitem{Maldacena98}
J. M. Maldacena, Adv. Theor. Math. Phys. {\bf 2}, 231 (1998).

\bibitem{Gubser98}
S. S. Gubser, I. R. Klebanov and A. M. Polyakov, Phys. Lett. B {\bf 428}, 105 (1998).

\bibitem{tHooft93}
G. \textquoteright t Hooft, in: {\it Salamfestschrift} (World Scientific, 1993) 0284-296; arXiv:gr-qc/9310026

\bibitem{Susskind95}
L. Susskind, J. Math. Phys. {\bf 36}, 6377 (1995).

\bibitem{RyuTakayanagi06a}
S. Ryu and T. Takayanagi, Phys. Rev. Lett. {\bf 96}, 181602 (2006).

\bibitem{RyuTakayanagi06b}
S. Ryu and T. Takayanagi, JHEP {\bf 0608}, 045 (2006).

\bibitem{Casini11}
H. Casini, M. Huerta, and R. C. Myers, JHEP {\bf 05}, 036 (2011). 

\bibitem{NielsenChuang}
M. A. Nielsen and I. L. Chuang, {\it Quantum Computation and Quantum Information} (Cambridge Univ. Press, 2011).

\bibitem{RieraLatorre06}
A. Riera and J. I. Latorre, Phys. Rev. A {\bf 74}, 052326 (2006).

\bibitem{Eisert10}
J. Eisert, M. Cramer, and M. B. Plenio, Rev. Mod. Phys. {\bf 82}, 277 (2010).

\bibitem{KitaevPreskill06}
A. Kitaev and K. Preskill, Phys. Rev. Lett. {\bf 96}, 110404 (2006).

\bibitem{Vidal07}
G. Vidal, Phys. Rev. Lett. {\bf 99}, 220405 (2007).

\bibitem{Haegeman13}
J. Haegeman, T. J. Osborne, H. Verschelde, and F. Verstraete,
Phys. Rev. Lett. {\bf 110}, 100402 (2013).

\bibitem{Swingle12}
B. Swingle, Phys. Rev. D {\bf 86}, 065007 (2012).

\bibitem{Eisler10}
V. Eisler and I. Peschel, Ann. Phys. (Berlin) {\bf 522}, 679 (2010).

\bibitem{Nozaki12}
M. Nozaki, S. Ryu, and T. Takayanagi, JHEP {\bf 10}, 193 (2012).

\bibitem{Wegner94}
F. Wegner, Ann. Phys. (Leipzig) {\bf 506}, 77 (1994).

\bibitem{Kehrein06}
S. Kehrein, {\it The  flow equation approach to many-particle systems} (Springer, Berlin, 2006).

\bibitem{Kehrein97}
S. Kehrein and A. Mielke, Ann. Physik (Leipzig) {\bf 6}, 90 (1997).

\bibitem{Kehrein99}
S. Kehrein, Phys. Rev. Lett. {\bf 83}, 4914 (1999).

\bibitem{footnote_uB}
In the flow equation method the flow is usually expressed with respect to a 
parameter~$B$ with dimension (Energy)${}^{-2}$ \cite{Kehrein06}.
The identification $B=\Lambda^{-2} e^{2u}$ connects the respective expressions.
The choice of a logarithmic flow parameter~$u$ turns out to be particularly useful for
the purposes of this Letter. 

\bibitem{Holzhey94}
C. Holzhey, F. Larsen, and F. Wilczek, Nucl. Phys. B {\bf 424}, 44 (1994).

\bibitem{Calabrese04}
P. Calabrese and J. Cardy, J. Stat. Mech. P06002 (2004).

\bibitem{Calabrese09}
P. Calabrese and J. Cardy, J. Phys. A {\bf 42}, 504005 (2009). 

\bibitem{footnote_check}
At half filling there is an exact analytical result for arbitrary~$g$ \cite{Eisler10}, which is plotted as the
dashed line in Fig.~\ref{fig1d}. It serves as a test for my exact diagonalization routine; its results agree within
numerical precision as can be seen in Fig.~\ref{fig1d}. 

\bibitem{footnote_ED2d}
For the translation-invariant case $g=1$ there is an analytic result \cite{Calabrese12} which has been used
to validate my exact diagonalization numerics for~$d=2$. The agreement is excellent for all~$\mu$ in Fig.~\ref{fig2d}
such that the differences are not visible on the scale of the plot. 

\bibitem{Calabrese12}
P. Calabrese, M. Mintchev, and E. Vicari, Europhys. Lett. {\bf 97}, 20009 (2012).

\bibitem{footnote_disjoint}
These considerations can be extended to the situation when the region~$B$ consists of multiple disjoint intervals. 
This will be worked out in a latter publication. 

\end{thebibliography}
\end{document}